\documentstyle[11pt,Cargesepasp,twoside,epsf]{article}
\def\edcomment#1{\iffalse\marginpar{\raggedright\sl#1\/}\else\relax\fi}
\reversemarginpar

\begin{document}
\title{Embedded Star Clusters: The ISO View}

\author{Anlaug Amanda Kaas}
\affil{ESTEC, Astrophysics Division, Noordwijk, The Netherlands \\
      Nordic Optical Telescope, Apdo 474, E-38700 Santa Cruz de La Palma}
\author{Sylvain Bontemps}
\affil{Observatoire de Bordeaux, BP 89, FR-33270 Floirac}

\begin{abstract}
We summarize the main results of a mid-IR photometric survey with ISO for a few
nearby embedded clusters. The sensitivity and spatial resolution of ISOCAM provided 
a wonderful tool for studies of the youngest stellar clusters, which are still 
deeply embedded in their parent molecular clouds. 
Mid-IR photometry is found to be extremely efficient in finding all the young 
stellar objects (YSOs) with IR excesses, i.e. mainly T~Tauri stars surrounded by 
circumstellar disks and also some Class\,I sources. The spatial distribution of
these sources is highly clustered and even sub-clustered, with a tendency of
stronger concentration for the younger regions.
The improved samples of IR-excess YSOs appear complete down to $0.03\,L_\odot$ for 
the most nearby clusters. This gives constraints on the mass functions well into 
the brown dwarf domain. The first results show that the mass functions of these 
clusters are identical to the field star IMF derived by Kroupa et al. (1993) 
with a flattening of the mass function at $M_\star \sim 0.5\,M_\odot$.
\end{abstract}

\section{Introduction}

Large scale cluster properties, such as the stellar Initial Mass Function (IMF) 
and the spatial distributions, are likely to hold important clues to the questions
related to the origin and early evolution of stellar clusters. In order to address 
these problems we approach the large scale properties observationally in as young 
clusters as possible. By doing that, we should avoid the potential incompleteness 
problems often encountered in studies of clusters of ages $> 10^8$ yrs, such as mass 
segregation and loss of low-mass members through dynamical evolution. In addition,
in these young regions there is a closer link between the stars and the material from 
which the stars form, such that studies of the cloud morphology and e.g. the clump mass 
functions can be directly compared to the stellar mass functions. Detection and 
classification of stellar and sub-stellar members of embedded clusters require 
sensitivity and spatial resolution in the infrared (IR). Since the majority of the 
cluster members are contracting down the Hayashi track, or in some cases even in 
their protostellar phases at these young ages, they are more luminous than later on, 
especially in the IR. This paper reviews what has been discerned about embedded star 
clusters from ISOCAM, the mid-IR camera aboard the Infrared Space Observatory (Cesarsky 
et al. 1996; Kessler et al. 1996).

Before ISOCAM, our knowledge of the stellar content of embedded clusters was in most 
cases limited to observations in the near-IR and the IRAS survey. Extensive ground
based surveys in the J (1.25 $\mu$m), H (1.65 $\mu$m), and K (2.2 $\mu$m) bands have 
revealed numerous Young Stellar Objects (YSOs) clustered in star formation regions 
(e.g. Eiroa \& Casali 1992, Prusti et al. 1992; Wilking et al. 1997; Barsony et al. 1997), 
while mid-IR ground based data have been more scarce (Lada \& Wilking 1984, 
Greene et al. 1994), and the spatial resolution of IRAS has been a limitation.  
See Wilking (2000) for an observational review on embedded clusters. 

Section~2 gives an overview of the ISOCAM star formation surveys, Section~3 presents the
general result of improved YSO samples based on the IR excess criterion, Section~4 discusses 
the nature of the IR-excess population, Section~5 the spatial distributions, Section~6 the 
improved luminosity functions, and Section~7 the implications for the IMF.

\section{The ISOCAM surveys of star-formation regions}

The surveys LNORDH.SURVEY\_1 and GOLOFSSO.D\_SURMC within the ISO central programme 
mapped selected parts of the major nearby star formation regions in the two broad band 
filters LW2 (5-8.5 $\mu$m) and LW3 (12-18 $\mu$m), designed to avoid the silicate features, 
and selected to sample the mid-IR Spectral Energy Distribution (SED) of YSOs in different 
evolutionary phases. The objective of the survey was to make a deep search for low-mass 
YSOs.

A review of these surveys was given by Nordh et al. (1998). The first results increased
the number of recognized YSOs in the Chamaeleon dark clouds by almost a factor of 2  
(Nordh et al. 1996). Olofsson et al. (1998) obtained a luminosity function for the 
Chamaeleon I IR-excess YSOs, finding that a substantial fraction of these objects could 
be young brown dwarfs with circumstellar disks, as suggested from their IR colours. An 
improved luminosity function from re-analyzed data was presented by Persi et al. (2000). 
The first results from the R Corona Australis region are found in Olofsson et al. (1999). 
The $\rho$ Ophiuchi embedded cluster is presented by Bontemps et al. (2000a, 2000b) 
with a thorough derivation of the luminosity and mass function. ISOCAM results on the 
Serpens Cloud are found in Kaas et al. (1999, 2000a, 2000b). Analysis of other regions 
is on-going, and specific results will be detailed elsewhere.

\section{Deep sample of embedded YSOs with ISOCAM}

The observational challenge in a statistical study of an embedded cluster is primarily 
to detect the stellar objects, then to separate the cluster members from field stars, 
and finally to determine intrinsic properties such as luminosity and mass of the 
members. The degree of embeddedness (A$_V \sim$ 10-100) requires sensitive IR 
observations, and the clustering requires good spatial resolution. With the 
3-6$\arcsec$/pix resolution of ISOCAM, and with an achievable sensitivity of a few mJy 
for $\sim$ 15 s on-source integration at 6.7 and 14.3 $\mu$m, we are able to sample 
objects down to at least 0.02 $L_{\odot}$ for the nearby regions. This corresponds to 
masses well into the brown dwarf regime, for reasonable estimates of the ages.

A summary of the main observational result in Chamaeleon, RCrA, $\rho$ Ophiuchi, and
Serpens is given in Fig.~1a)-d), where we plot the colour index $[14.3/6.7]$, defined as 
the logarithm of the flux ratio
between the two filters, versus the flux at 14.3 $\mu$m (in Jansky). As an alternative to
the $[14.3/6.7]$ index, one can calculate the SED index $\alpha = d \log (\lambda 
F_{\lambda})/ d \log \lambda $ between 6.7 and 14.3 $\mu$m, shown on the right-hand y-axes. 
In each region the sources fall in two separate groups. One is located at $ [14.3/6.7] 
\approx -0.66$ ($\alpha_{6.7-14.3} \approx -3$), which is the colour expected for normal 
photospheres, i.e. the slope in the Rayleigh-Jeans tail. The spread around this value 
mainly indicates the increasing photometric uncertainty with decreasing flux. The effect 
of extinction is small, and a reddening vector of size corresponding to A$_K = 3$ 
(A$_V \sim$ 30) is indicated in each panel. The second group of sources comprises 
all objects located well above the first group. These are sources with mid-IR excess 
emission. Most of these sources have colours typical of classical T Tauri stars (CTTS), 
and the IR excess is interpreted in terms of optically thick circumstellar disk emission. 

\begin{figure}
\plottwo{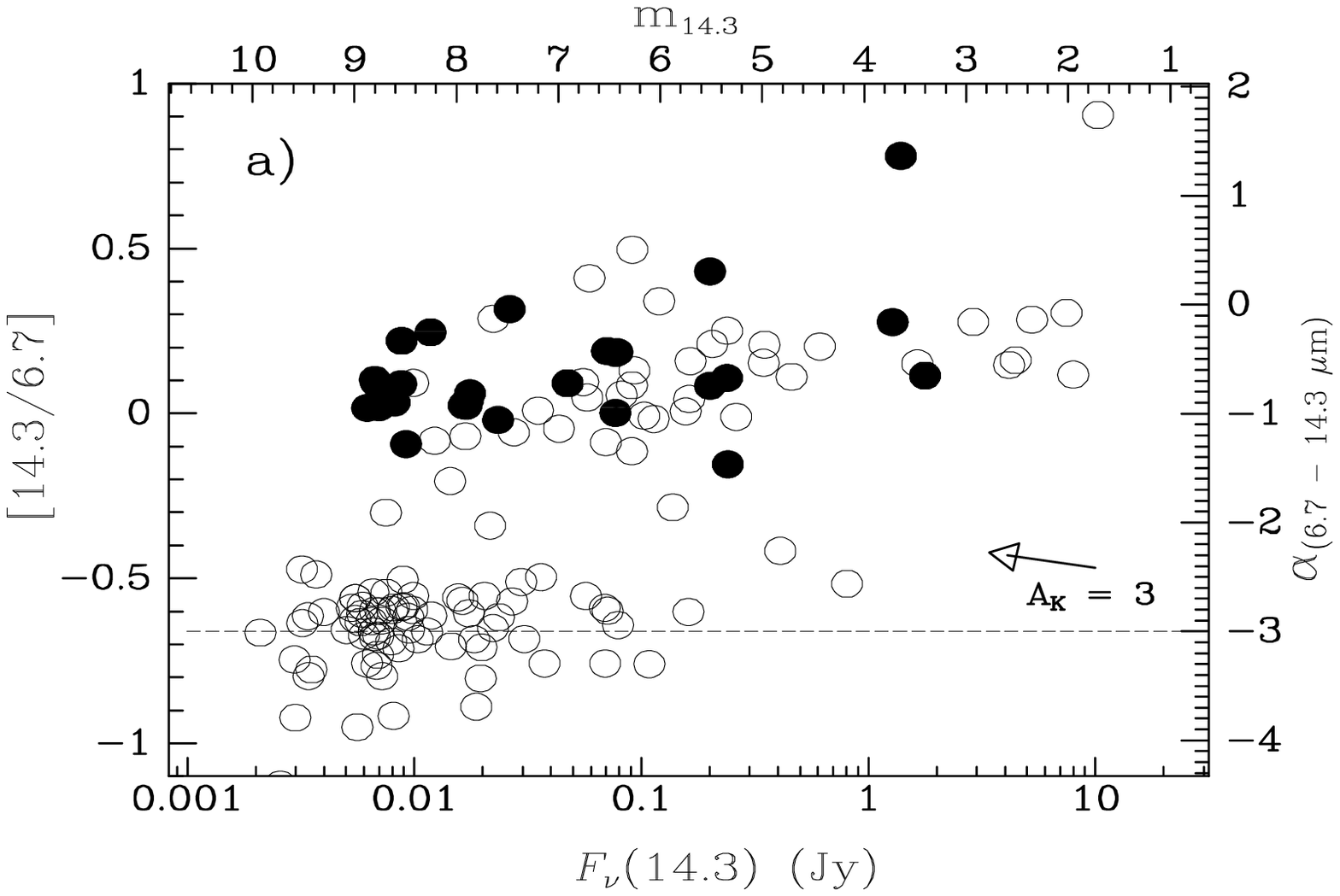}{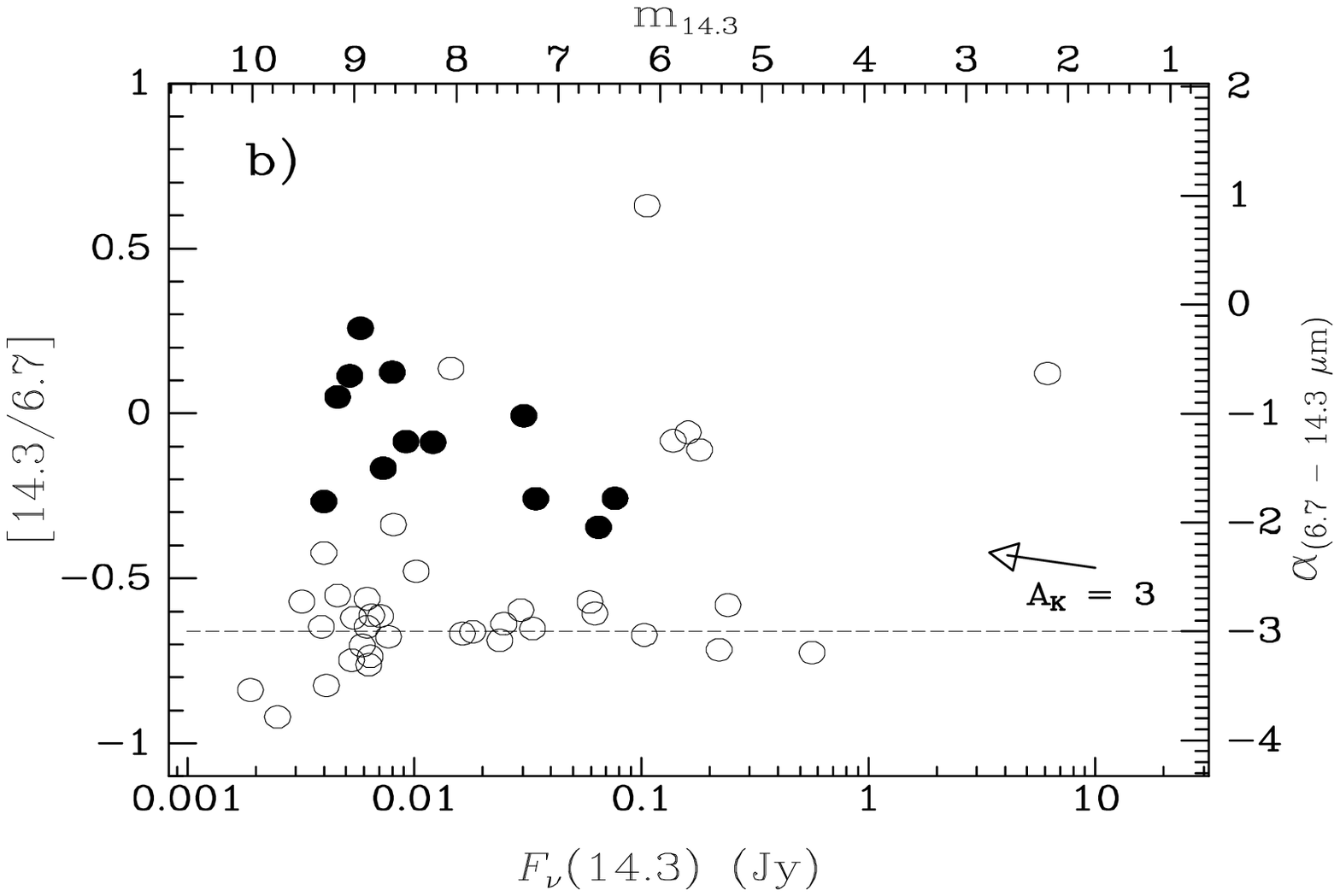}
\plottwo{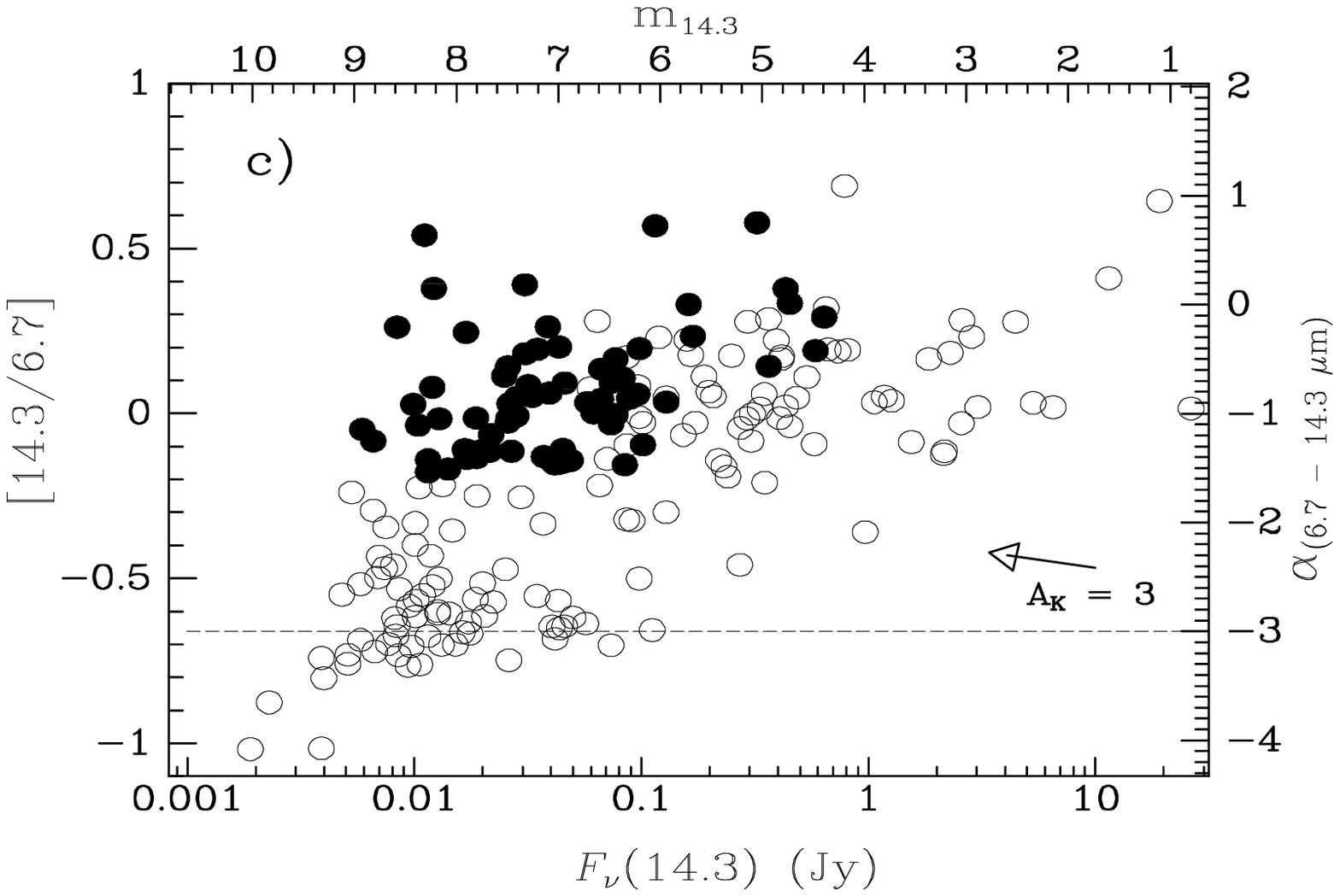}{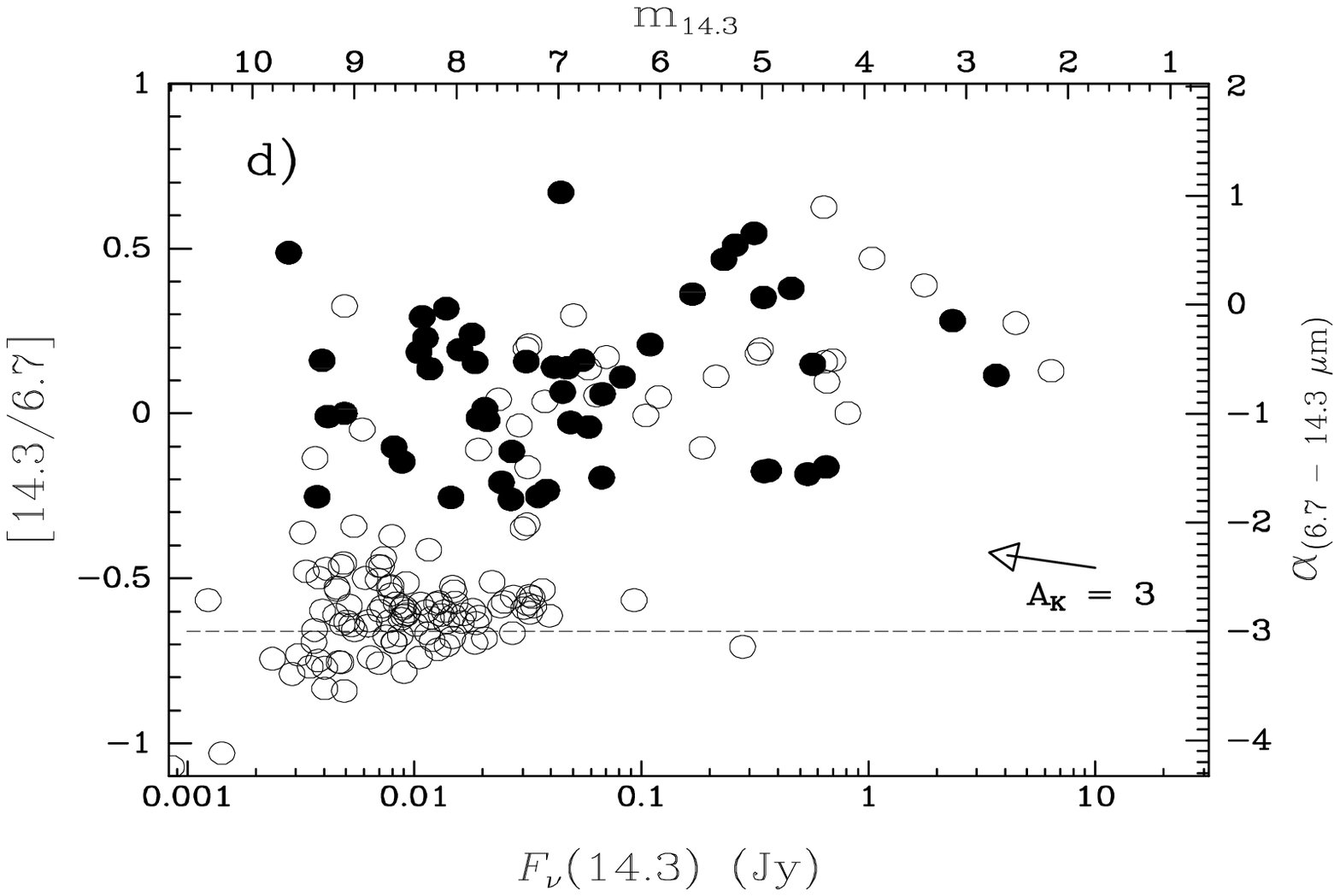}
\caption{The color index $[14.3/6.7]$ vs. the 14.3 $\mu$m flux (Jy) for a) Chamaeleon I,
         II, and III, b) R Corona Australis, c) $\rho$ Ophiuchi and d) Serpens. New YSOs
         are marked as filled circles. }
\end{figure}

Fig.~1~a) shows sources from 3 maps (about 0.8 square degrees) located in Cha I 
(Persi et al. 2000), Cha II and Cha III (Nordh et al. 1996). Fig.~1~b) shows point 
sources from $\sim$ 0.42 square degrees in RCrA (Olofsson et al. 1999), where the 
central Coronet cluster could not be mapped due to detector saturation precautions. 
The $\rho$ Ophiuchi sample from a total of 0.7 square degrees mapping (Bontemps et al. 
2000a,b) is shown in Fig.~1~c). Fig.~1~d) shows the sample from a 0.22 square degree 
survey of the Serpens Cloud (Kaas et al. 1999). Filled circles indicate the new 
YSOs found with ISOCAM.

Practically all sources detected with ISOCAM have near-IR K-band counterparts; although
in dense clouds and for the youngest objects, relatively deep K-band ($>$ 15 mag) data 
might be required. Thus, rather than having discovered previously unseen objects, the
main value of the ISOCAM surveys has been to provide a means by which IR-excess YSOs 
can be easily distinguished from field stars, mainly due to the small effect of reddening 
in the mid-IR. YSOs without IR-excesses, on the other hand, can not be recognized, and 
alternative approaches, such as X-ray surveys (Feigelson \& Montmerle 1999), appear 
promising in order to sample this part of the cluster population. ISOCAM has, on the 
average, doubled the IR-excess populations.

\begin{figure}
\plottwo{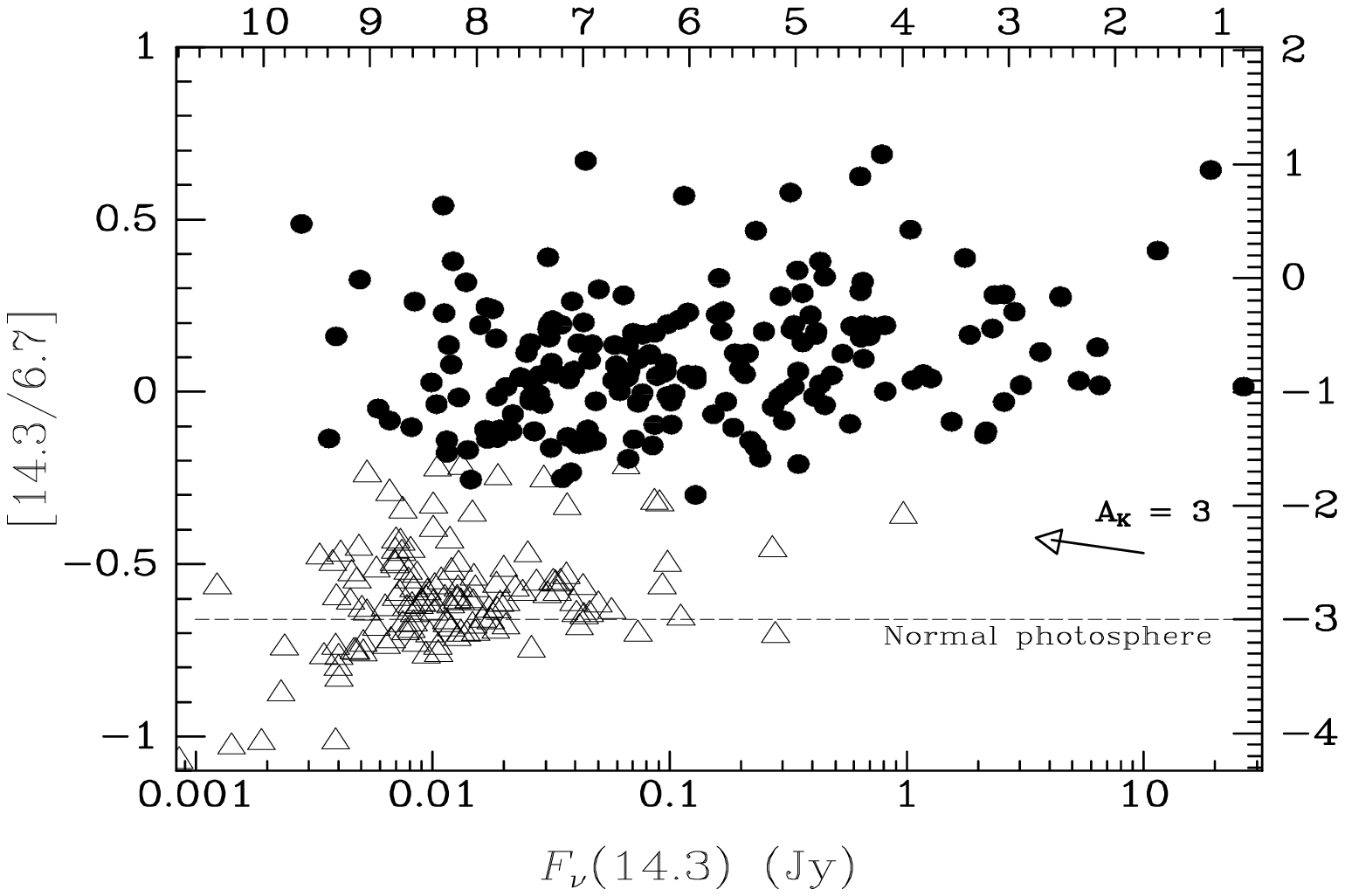}{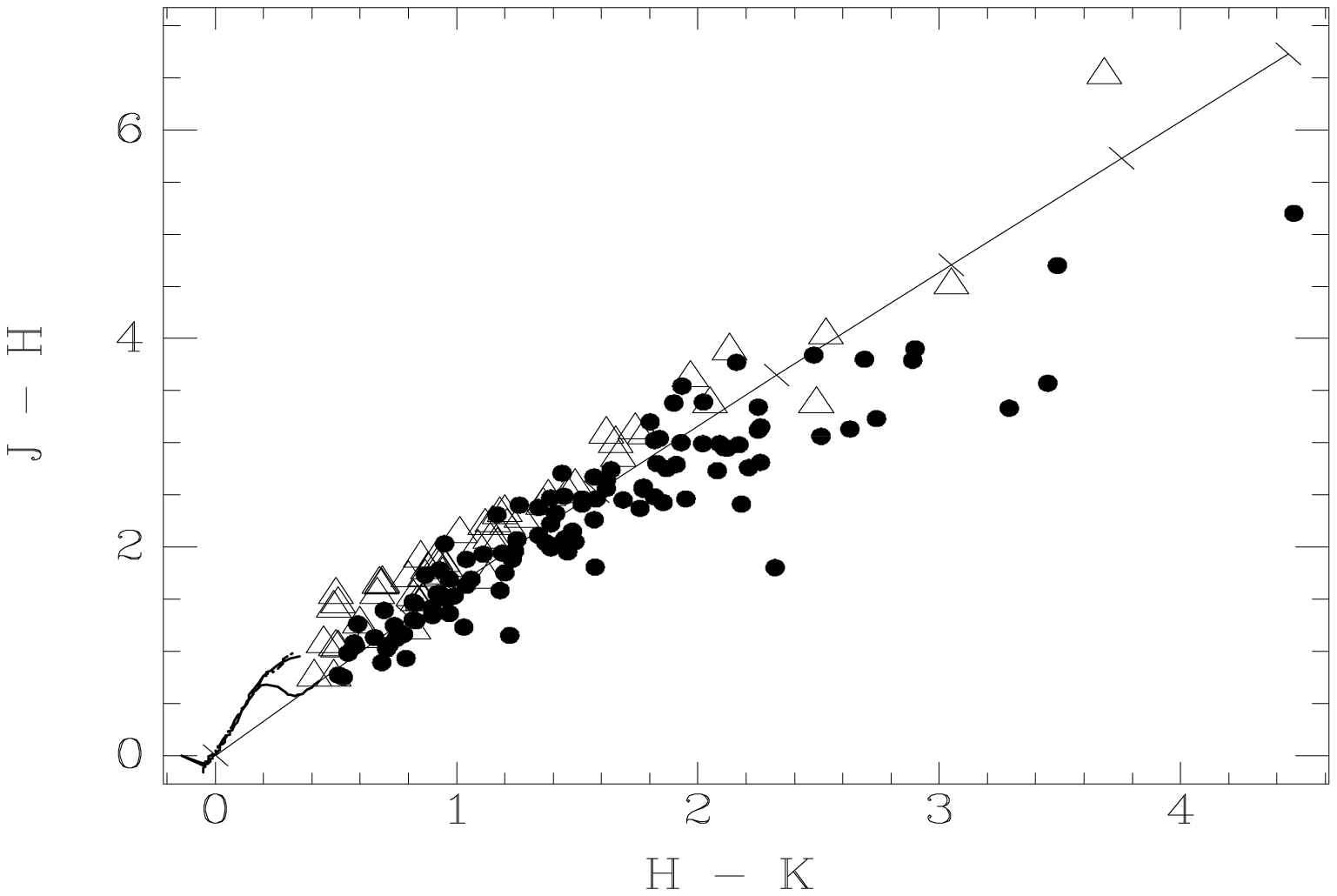}
\caption{Comparison of the single [14.3/6.7] colour (left) with the J-H/H-K diagram 
         (right) for the same sources. About 50 \% of the IR-excess sources (filled
         circles) are found sufficiently to the right of the reddening line (slope
         is 1.56) in the $J-H/H-K$ diagram.}
\end{figure}

While circumstellar dust easily is recognized by the IR excesses that show up in broad 
band photometry, it is necessary, however, to go to sufficiently long wavelengths in order 
to clearly separate the effects of cloud extinction to those of intrinsic IR excess. This 
is illustrated in Fig.~2 where we compare the frequently used $J-H/H-K$ colour diagram with 
the single colour index $[14.3/6.7]$ for sources from $\rho$~Ophiuchi and Serpens with 
available ISOCAM and JHK photometry. The left panel shows a clear separation of the IR 
excess population (filled circles). In the $J-H/H-K$ diagram, however, sources with 
IR-excess (filled circles) mix with those without IR excesses (open triangles) and 
cluster along the reddening vector. We find that on the average about 50\% of the mid-IR 
excess population have near-IR excesses in the $J-H/H-K$ diagram (e.g. Kaas et al. 2000b). 
Thus, studies of embedded clusters based on JHK data alone are seriously incomplete, and 
any determination of disk frequencies based on such studies must be interpreted with care.

The IR excesses, so apparent in the $[14.3/6.7]$ colour index, show up already at 
6.7 $\mu$m, and both the $J-K/K-m_{6.7}$ diagram (Olofsson et al. 1999, Persi et al. 2000) 
and the $H-K/K-m_{6.7}$ diagram (Kaas et al. 2000b) are efficient in distinguishing 
between intrinsic IR excess and effects of scattering and extinction. These diagrams are used 
to sample faint IR-excess sources not detected at 14.3 $\mu$m. 

\section{The nature of the IR-excess sources}

According to the current empirical picture of early stellar evolution (Adams et al. 
1987; Lada 1987; Andr\'e et al. 1993), young stars can be observationally classified 
along the evolutionary sequence: Class\,0 - Class\,I - Class\,II - Class\,III; 
corresponding to a gradual loss of circumstellar matter with age. For sources with 
near-IR and mid-IR photometry, one can use the classical parametrization of the Spectral 
Energy Distribution (SED) and calculate the mid-IR spectral index $\alpha = d \log (\lambda 
F_{\lambda})/ d \log \lambda $, usually taken between 2 and 10 $\mu$m (Wilking \& Lada 1984). 
In Fig.~3 we plot $\alpha_{2.2-6.7}$ versus $\alpha_{2.2-14.3}$, or equivalently 
the $K-m_{6.7}/K-m_{14.3}$ colour diagram, for a subset of our sources and draw the YSO class 
loci according to current definitions. Previously known Class\,I sources (large stars), 
Class\,II sources (filled circles), and Class\,III sources (filled triangles) appear mainly 
within their respective loci.

\begin{figure}
\plotone{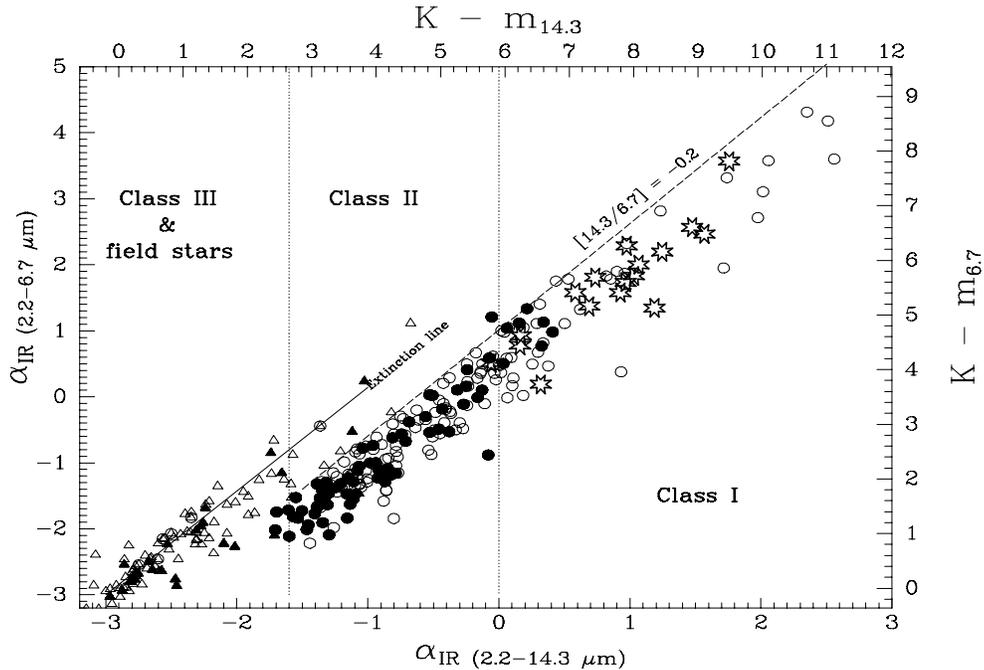}
\caption{All sources with ISOCAM and K-band photometry are plotted as $\alpha_{2.2-6.7}$
         vs. $\alpha_{2.2-14.3}$ or equivalently: $K-m_{6.7}/K-m_{14.3}$.}
\end{figure}

Following this simplified IR spectral classification scheme, the new IR-excess population 
(open circles) can now be tentatively classified as Class\,I and Class\,II sources.  
In Chamaeleon~I and in $\rho$~Ophiuchi the resulting Class\,I/Class\,II number ratio is about 
$0.1-0.2$, as expected based on current ideas about the timescales for each phase. In the Serpens 
Cloud Core, however, this ratio is appreciably larger (close to 1), and this has been 
interpreted in terms of an extremely young age for this region (Kaas et al. 2000a, 2000b).

The strong influence of extinction at 2.2 $\mu$m is apparent. Sources without IR-excesses 
(triangles) are outlining the reddening line in this diagram. These are a mix of reddened 
field stars and Class\,III sources, and the large scatter is probably from the contribution 
of small IR excesses in a number of Class\,II - Class\,III transition objects. The border 
used in Fig.~1 to separate between Class\,II and Class\,III sources, at $[14.3/6.7] = -0.2$ 
corresponding to $\alpha_{6.7-14.3} = -1.6$, is shown as a dashed slope. The mid-IR excess 
group to the right follows roughly the same slope as the reddening vector, and Class\,I 
sources are found at the extreme end of the $K-m_{14.3}$ colour, rather than at the extreme 
of the $[14.3/6.7]$ colour. We speculate that this roughly similar $[14.3/6.7]$ colour for 
Class\,II and Class\,I sources could be due to dust properties such as broad silicate absorption 
features or the presence of H$_2$O and CO$_2$ ices in the 14.3 $\mu$m band for the more
embedded sources (Class\,Is), but this requires further study.

Since we cannot distinguish between Class\,III sources and field stars from broad band 
photometry alone, we can only give rough upper limits on the disk frequencies in the 
populations sampled by ISOCAM, based on previous studies of the stellar populations and/or
models of the mid-IR point source sky. As an example, we mention that within the region 
covered by ISOCAM in Chamaeleon I, the fraction of IR-excess sources among currently 
known YSOs is 76\% (Persi et al. 2000). There could be more Class\,IIIs, however, among 
the non-IR-excess sources. Also, the fraction of disk sources among YSOs should be 
counted as a function of cluster radius or cloud density, in order to compare between 
the regions.

\section{Spatial distributions}

The spatial distribution of the IR-excess YSOs is always found clustered at the scale of the 
molecular clouds, but also more locally, as a few sub-clusters which are normally associated 
with the dense cores of the molecular clouds. In Chamaeleon~I clustering is apparent around 
the reflection nebula Ced 112 on a size scale of about 10-15 arcmin or 0.5-0.7 pc, while 
the clustering around Ced 110 and Ced 111 is less confined (Nordh et al. 1996; Persi et al. 
2000). In $\rho$~Ophiuchi 4 sub-clusters are found associated with dense cores: Oph A, B, 
EF (between Oph E and Oph F) and L1689S. These have approximate sizes of 4-8 arcmin, or
0.2-0.3 pc (Bontemps et al. 2000b). For comparison, the spatial distribution in TMC-1 shows no 
sign of clustering.

\begin{figure}
\plotfiddle{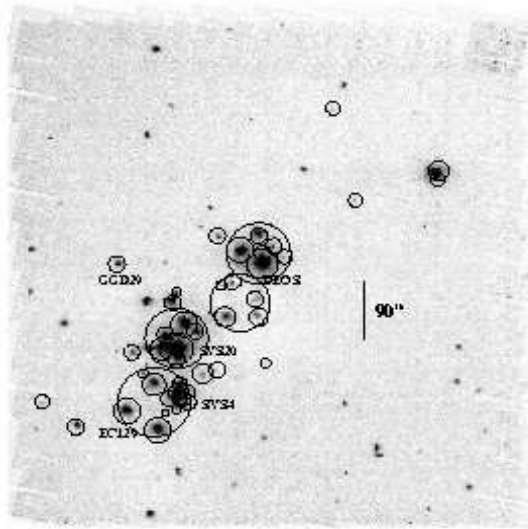}{7.5cm}{0}{100}{95}{-130}{0}
\caption{ISOCAM image of 13' $\times$ 13' covering the Serpens Cloud Core. Encircled sources
are of Class\,II or Class\,I type, and these are mainly found in 4 sub-clusters (large circles)
along a NW-SE oriented ridge, spatially corresponding to 4 kinematically separated N$_2$H$^+$ 
cores A,B,C,D (Testi et al. 2000).}
\end{figure}

Figure~4 shows a 13' $\times$ 13' ISOCAM image covering the Serpens Cloud Core ($\sim$ 1 
square pc for a distance of 260 pc). The YSOs with sufficient IR-excess to be of Class\,II 
and Class\,I type (encircled) mainly follow a NW-SE oriented ridge, and most are further 
seen clustered into 4 dense sub-clusters (large circles), each with a diameter of about 
1.5 arcmin, or $\sim$ 0.1 pc. Each of these contains between 5 and 14 YSOs in general (at 
the sensitivity and resolution of ISOCAM), and between 2 and 6 Class\,I sources in 
particular (Kaas et al. 2000a,b). The unusually rich collection of clustered Class\,I sources 
strongly argues in favour of this region being especially young. 
The locations of the sub-clusters coincide well with the distribution of tracers of the 
dense cloud, such as 1100 $\mu$m continuum emission (Casali et al. 1993), C$^{18}$O line 
emission (White et al. 1995), and high resolution 3 mm continuum emission (Testi \& Sargent 
1998). In particular, the 4 clusters coincide with the spatially and kinematically separated 
N$_2$H$^+$ cores A,B,C,D found by Testi et al. (2000), which strongly supports their claim 
of an observational evidence for hierachical fragmentation within a cluster-forming core. 

\section{Luminosity and mass functions}

As shown in the previous sections, mid-IR surveys significantly improve the census of 
young stars in embedded clusters. This provides new and larger samples of newly formed 
stars in these star forming regions. The resulting luminosity functions (LF) are therefore
more complete and give more precise constraints on the distribution of stellar masses 
which can give clues to the origin of the galactic IMF. The embedded young stars are, 
however, difficult to study, and even their stellar luminosities are not easy to estimate 
from simple photometric data (without spectroscopy). 

For embedded (not optically visible) pre-main sequence stars, the most reliable method 
to estimate stellar luminosities from photometric data consists in deriving dereddened I 
or J band magnitudes and to convert them to stellar luminosities. This is possible 
because I and J are close to the maximum of the photospheric energy distribution and not 
too much affected by intrinsic IR excesses (e.g. Greene et al. 1994, Persi et al. 2000, 
Bontemps et al. 2000b). The resulting luminosity functions for the $\rho$~Ophiuchi (140~pc) 
and the Chamaeleon~I (160~pc) clusters are shown in Fig.~5. These luminosity functions are 
actually limited to Class~II YSOs. The Class~III YSO samples are still too incomplete to be 
included.

\begin{figure}
\plotone{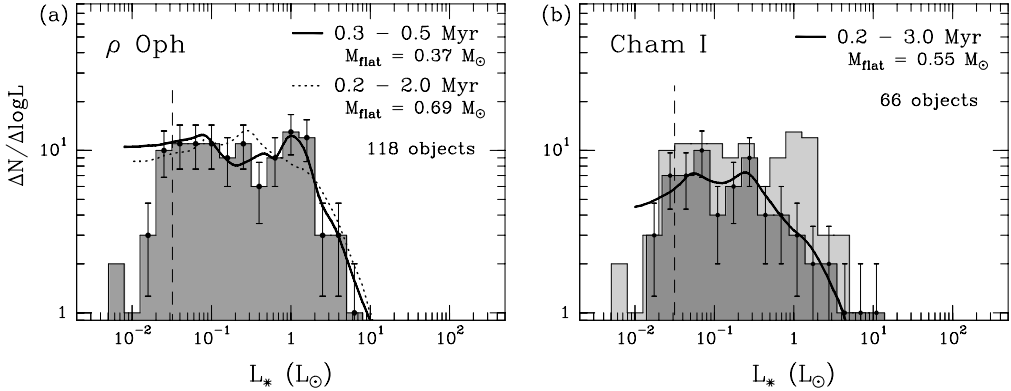}
\caption{The stellar luminosity functions for Class~II YSOs (embedded CTTS) as derived 
         from near-IR data for the two following clusters: (a) $\rho$~Ophiuchi (140~pc); 
         (b) Chamealeon I (160~pc). In (b) the $\rho$~Ophiuchi luminosity function is 
         displayed in the background for comparison. The dashed vertical lines mark the 
         estimated completeness levels in $L_\star$. The curves display the results of a 
         simple modeling from Bontemps et al. (2000b), (see text for details).}
\end{figure}

The global shapes of these two luminosity functions are similar: the number of stars per 
logarithmic $L_\star$ bin sharply decreases for $L_\star> 0.5 - 1.0\,L_\odot$ and is 
more or less constant at low luminosities, down to the completeness level, which is equal 
to $\sim 0.03\,L_\odot$ in both cases. However, the flattening occurs at $L_\star \sim 
1.5\,L_\odot$ for $\rho$~Ophiuchi while at $\sim 0.3\,L_\odot$ for Chamaeleon~I. 
Even if the ``flattening luminosity'' for Chamaeleon~I is rather uncertain due to the smaller 
number of stars (66 instead of 118 for $\rho$~Ophiuchi), the $\rho$~Ophiuchi luminosity 
function certainly flattens out at a higher luminosity than the Chamaeleon~I LF. According to 
Bontemps et al. (2000b), the clear flattening over two orders of magnitude in the 
$\rho$~Ophiuchi cluster can be explained only by a similar flattening of the underlying 
mass function. Assuming a flat (in logarithmic units) mass function, these authors
have built a simple modeling of the LFs using D'Antona \& Mazzitelli (1998) pre-main sequence 
tracks and assuming some simple age distributions for the young stellar population (see 
Bontemps et al. 2000b for details). The results of this modeling are shown in Fig.~5 for 
$\rho$~Ophiuchi and Chamaeleon~I. The parameters for these fits (made on the parameter 
$M_{\mathrm flat}$ at which the mass function flattens out) are given in the plots. 
For $\rho$~Ophiuchi the best fit is obtained with the combination of a rather small age of 
0.4~Myr for the Class~IIs and $M_{\mathrm flat} = 0.37\,M_\odot$. A model with a broad 
distribution of ages between 0.2 and 2.0~Myr, and a larger flattening mass $M_{\mathrm flat} 
= 0.69\,M_\odot$ can, however, also fit the data acceptably. A small age for the $\rho$~Ophiuchi 
embedded population has been favored by recent, independent works (Greene \& Meyer 1995; 
Luhman \& Rieke 1999). Our new results might therefore support very well these previous 
suggestions. Fig.~5b shows the result of a similar modeling for the Chamaeleon~I data 
(reduced to the 66 Class~IIs with $L_\star$ estimates in Persi et al. 2000). It shows that a 
good modeling of the observed luminosity function is also possible here. An older population 
of stars is used because there are good evidences that the Chamaeleon~I region is older than 
$\rho$~Ophiuchi (e.g. Lawson et al. 1996). With such a larger average age, the flattening at 
a lower luminosity can be reproduced with more or less the same mass function as for 
$\rho$~Ophiuchi: the best fit is obtained with $M_{\mathrm flat}= 0.55\,M_\odot$.

\section{Implications for the IMF}

The constraints on the mass functions within the surveyed clusters, which are born in 
typical star-forming regions, are expected to ultimately explain the origin of the field 
star IMF. Up to now a mass function has been derived only for the $\rho$~Ophiuchi cluster. 
After correction for hidden binaries, Bontemps et al. (2000b) obtained a single star mass 
function which is very similar to the field star IMF derived by Kroupa et al. (1993). 
No turnover or local maximum is recognized down to $0.05\,M_\odot$. This shows that a 
large number of young brown dwarfs are actually present in the $\rho$~Ophiuchi cluster. 
Thirty brown-dwarf candidates\footnote{They all have a mid-IR excess since they have been 
selected using this criterium.} have been detected there.

\section{Conclusions}

The ISOCAM survey has significantly improved our understanding of embedded clusters. Because
the samples of IR-excess YSOs are now much more complete, the IMF can be studied with confidence
well into the brown dwarf domain for the nearest regions, and the distribution of the youngest 
YSOs (Class\,I and Class\,II sources) can be studied on small spatial scales. There is good 
correspondance between the currently estimated ages of the 3 star formation regions Chamaeleon~I, 
$\rho$~Ophiuch and Serpens, and the size of their sub-clusters of IR-excess YSOs. 
We conclude that the luminosity functions which are complete down to $\sim 0.03\,L_\odot$ 
can easily be modeled with plausible age distributions. The mass functions needed to 
reproduce the observations are roughly flat in logarithmic units in the low-mass range, 
and the flattening occurs at $M_\star$ of the order of $0.5\,M_\odot$. For future work 
we would like to build luminosity functions for more clusters to test whether such a flat 
mass function can always reproduce the observations.



\end{document}